# Water Solvation of Charged and Neutral Gold Nanoparticles

Fabio Novelli[1], Miguel Bernal Lopez[2], Gerhard Schwaab[1], Beatriz Roldan Cuenya[2,3], Martina Havenith[1,*]

[1]Department of Physical Chemistry II, Ruhr University Bochum, 44780 Bochum, Germany

[2]Department of Physics, Ruhr University Bochum, 44780 Bochum, Germany

[3]Department of Interface Science, Fritz-Haber Institute of the Max Planck Society, Berlin 14195, Germany

*martina.havenith@rub.de

## Abstract

*Gold nanoparticles are unique electrocatalysts for oxygen reduction, carbon dioxide reduction, and alcohol oxidation. Electrocatalytic processes are influenced by the interaction with the solvent, yet the direct investigation of the solvation of nanoparticles is scarce. Here we select gold nanoparticles as ~10 nm sized solutes of which we can control the charge. We perform mid-infrared and terahertz spectroscopy to compare the behavior of the water in solution with micelles loaded with neutral and positive gold nanoparticles. We find indications that the hydration water around the gold-loaded micelles is characterized by weaker hydrogen bonds than bulk water. A positive nanoparticle charge is observed to result in a larger blue shift of the OH stretch, quenches the intensity of the collective translational mode, and increases the absorption by librating water molecules. Water at the interface of a positive gold nanoparticle could experience a stiffer potential energy surface which, in turn, might unveil local thermodynamic properties.*

## Introduction

Nanoparticles (NPs) may present higher catalytic activity compared to bulk materials due to their larger surface to volume ratio. Gold nanoparticle electrodes have peculiar electrocatalytic activity in reactions such as oxygen reduction reaction (ORR)[1–5], reduction of carbon dioxide ($CO_2$RR)[6–8], and alcohol oxidation[9–12]. The weak interaction between gold electrodes and CO or $CO_2$ makes gold a good catalyst for $CO_2$RR and alcohol oxidation. The prompt desorption of CO from the electrode surface during $CO_2$RR results in a high selectivity towards the formation of CO[13,14]. During the oxidation of alcohols, the desorption of CO avoids the poisoning of the surface[15]. For these reasons, gold NPs are attractive materials to be used as electrodes. The electrocatalytic activity is also determined by the interaction of the solvent used with the electrode and the electroactive species[16–18]. However, so far the experimental[19–21] characterization of the hydration properties of a gold NP is, to the best of our knowledge, lacking.

Gold-loaded micelles are suitable templates for the formation of well-defined and size-controlled NP electrodes[12,14,22–24]. Here we reveal how water molecules in solution with gold-loaded micelles are affected by the interaction with gold NPs. We performed mid-infrared (MIR) measurements, which are sensitive to intramolecular water vibrations[25–29], and terahertz (THz) spectroscopy, which probes directly the collective translational and librational motions of hydrogen-bonded water molecules[29,30]. The MIR absorption of water in solution with either a charged or a neutral NP is found to be blue shifted with respect to bulk water. The intermolecular hydrogen bond stretch of water in solution with the nanoparticles is red shifted with respect to the bulk. These observations are consistent with a weaker hydrogen bonding for the water molecules in solution with gold NPs. We speculate that water seems unable to form a complete 3D network around nanoparticles.

When comparing the response of water in solution with neutral and charged gold NPs, we found that a positively charged NP blue shifts the water absorption in the MIR. This suggests that a positively charged gold nanoparticle further weakens the hydrogen bonding of the water molecules. In the THz range, a positive gold NP enhances the intensity of the water librations (+32±1%) and quenches the intensity of intermolecular hydrogen bond stretch (-19±2%). The central frequencies and widths of these collective water modes are, however, the same in solution with charged or neutral gold NPs. We suggest that the unchanged central frequencies of the collective bands might be the result of a stiffer local potential[29,31] of the water at the interface with the charged gold nanoparticles.

**Experimental Methods.**

**Samples preparation**. In a typical inverse micelle, the head groups are at the core and encapsulate polar solvent molecules, while the hydrophobic ligands are the external tails and interact with another, non-polar solvent[32,33]. However, it also possible that diblock copolymers self-assembly into micelles in pure non-polar solvents[34]. Here, inverse micelles are formed by dissolving Poly(styrene-b-2-vinylpyridine) (PS-P2VP) diblock copolymers in water-insoluble pure toluene, or tetrahydrofuran (THF)[35–39]. The preparation of inverse micelles loaded with gold was conducted inside a glove box under controlled $N_2$ atmosphere. We obtain NPs made of charged gold atoms ($Au^{3+}$) by loading the micellar cages with ~20 mM $HAuCl_4$ and ~60 mM $H_2O$. In order to have neutral NP ($Au^0$), the $Au^{3+}$ species are reduced by addition of sodium borohydride with a $HAuCl_4$ to $NaBH_4$ molar ratio of 1:60. After the reduction of $Au^{3+}$ species, the micellar solution was filtered, removing the excess of solid $NaBH_4$. The infrared absorption of $NaBH_4$ could not be studied separately because of its very low solubility in non-protic solvents such as THF and toluene[40]. Given its low solubility and the large excess solid removed from the solution, in the following we neglect the contribution of $NaBH_4$.

**Characterization**. The color of the charged and neutral NP solutions are, respectively, light yellow and dark orange. This proves that the nanoparticles have different charge. For $Au^{3+}$ NP, the plasma frequency is reduced and the real part of the dielectric function of the NP does not sustain the surface plasmon resonance (SPR) at ~560 nm[41–45] - see Supplementary Fig. 1.

The average NPs size is determined to be 5-20 nm by combined UV-VIS (Supplementary Fig. 1) and AFM measurements (Supplementary Fig. 2). Supplementary Fig. 2 shows Au nanoparticles supported on a Si wafer by dip coating[36]. In this method, the Si wafer substrate is immersed in the Au-loaded micellar solution and then pulled out, the solvent dries and the Au-loaded micelles are deposited. After the deposition, the surfactant polymer is removed with oxygen plasma. Supplementary Fig. 2 shows mono-dispersed Au nanoparticles that were inside Au-loaded micelles prior to the surfactant polymer removal. Fig.3 of ref[36] reports the intensity of the XPS spectra from the gold-loaded inverse micelles deposited on a Si substrate, as a function of the duration of the oxygen plasma treatment. The Au-4d features have approximately the same height, and thus there is about the same amount of gold, irrespective of how much polymer is removed. Thus, the AFM images shown in Supplementary Fig. 2 give an upper limit to the size of both neutral and charged NPs.

For the non-reduced, yellow $Au^{3+}$ solution, we previously imaged the $Au^{3+}$ NPs with TEM before and after polymer removal[38,39]. Before the polymer removal, $Au^{3+}$ ions form small clusters (~1-5 nm) inside the larger micellar cage, and in between those small gold clusters there were empty spaces in the core (see e.g. Fig.1 in ref[38]). Once we remove the polymer, we see significantly consolidated NPs without the empty spaces with TEM. For the reduced, red $Au^0$ solution, the size of the NP in solution can be estimated from the SPR peak[41,42]. Unfortunately, a rigorous estimate requires to know the exact shape of the nanoparticle and the relative dielectric constant of the medium surrounding the NP, in our case the micelle core – neither of which are available (see e.g. eq.(1) in ref[42]). For this reason, we perform a qualitative comparison between the SPR measured in the $Au^0$-loaded micelles solutions with the ones from commercial solutions containing gold nano-spheres (Supplementary Fig. 1). The comparison suggest that the $Au^0$ NPs have a dimension comprised between about 10 nm and 100 nm.

As previously measured[35–39], all gold NPs are found at the core of the micelles, wrapped in the polar heads of the polymer. For example, see Fig.1 of ref[38] and Fig.3 of ref[37]. This is confirmed by dedicated DFT calculations[35], where the formation of the Au NPs inside the micellar nano-reactors was modeled by creating an encapsulating polymer head around the Au precursor units, and neglecting the solvent contribution. The calculations prove that the Au nanoparticle strongly bind to both N and C in the P2VP core of the reverse micelle, with a bond length of only[35] ~0.2 nm. Thus, the metal is confined inside the micelle, is wrapped by the polar heads of the core, and is ligand stabilized. Given the fact that the gold NPs are wrapped in the polymer[35], the inverse micelle interior is at least as large as the size of the NPs estimated after oxygen plasma removal of the ligand. For the non-reduced $Au^{3+}$ solution, we found indication[38,39] that we do not fill completely the entire micellar head, and the loading is about 0.6-0.7 metal salt molecular weight to P2VP (polymer head) ratio. As detailed in ref[38,39], the core of this kind of inverse micelles has been estimated previously to about 50 nm.

**Measurements**. The THz and MIR measurements reported in the manuscript (Figs. 1-3, SI Fig. 1, SI Fig. 3) are performed on gold-loaded or metal-free inverse micelles in solution with THF or toluene. A commercial Fourier transform infrared spectrometer was used to record spectra in the THz range between 50 $cm^{-1}$ and 700 $cm^{-1}$ (Fig. 1), and in the MIR within 3000 $cm^{-1}$ and 4000 $cm^{-1}$ (Fig. 2). We filled the liquid samples into a diamond cell kept at a temperature of 20±0.05 °C with a recirculating chiller. A spacer with nominal thickness of $d$=1 mm was used for the measurements of toluene solutions in the THz range. We chose $d$=0.21 mm for THz measurements on THF solutions. For MIR experiments we used a $d$=0.5 mm thick spacer. We measured the transmission of pure THF ($T^{THF}$) and toluene ($T^{tol}$) solvents, and of the corresponding solutions containing metal-free micelles ($T^{THF}_{mic}$ and $T^{tol}_{mic}$), micelles loaded with $Au^{3+}$ ($T^{THF}_{Au^{3+}}$ and $T^{tol}_{Au^{3+}}$), and $Au^0$ micelles ($T^{THF}_{Au^0}$ and $T^{tol}_{Au^0}$).

Water is miscible with THF but insoluble in toluene. From the concentration of the precursor, ~20 mM of ($HAuCl_4$ + 3x $H_2O$), water is present only in the gold-loaded micellar solutions in THF with a concentration of ~60 mM or[46–49] ~0.27 g water per g gold. We quantify the concentration of water by analyzing the MIR spectra shown in Supplementary Fig. 3. The water concentration is $c$=53 mM in solution with $Au^{3+}$ NP in THF, and $c$=63 mM in solution with $Au^0$ NP in THF. The dissolution of each $HAuCl_4$ forms[50] 1x $H^+$, 1x $Au^{3+}$, and 4x $Cl^-$. The average number of water molecules per solute is approximately 3x $H_2O$/6 = 0.5 $H_2O$/solute.

Considering a nanoparticle with a diameter of 10 nm, we estimate that this amount of water could form up to ~3 layers around each NP.

**Results**

**THz absorption**. We estimate the absorption coefficient of each solute $i$ (metal-free micelles, and micelles loaded with Au$^{3+}$ NP or Au$^0$ NP) to $\alpha_i^{solv} = -\frac{1}{d}\ln\frac{T_i^{solv}}{T^{solv}}$, where $d$ is the thickness of the spacer, $solv$ is the solvent used, and $T$ is the transmitted light intensity. In THF-based solutions, this absorption coefficient includes the contribution by water.

Fig. 1 displays the THz absorption coefficient of each solute in THF (blue) or toluene (black). The absorption of the metal-free micelles (Fig. 1a) is characterized by two peaks centered at 404 cm$^{-1}$ and 551 cm$^{-1}$. By comparison with previous works[51–53], we assign the 404 cm$^{-1}$ mode to a C-N-C ring vibration of the metal-free micelles (Fig. 1a). Fig. 1b (Fig. 1c) shows the THz absorption of the micelles loaded with charged (neutral) gold nanoparticles. Here, the C-N-C vibration at 404 cm$^{-1}$ vanishes and, instead, a sharp peak appears at 359 cm$^{-1}$. We note that Au-S vibrations are found at ~300 cm$^{-1}$ in thiolate-protected gold clusters[54]. DFT calculations on this micellar system[35] predict a strong bond between Au and N. Thus, we assign the strong mode at 359 cm$^{-1}$ to Au-N vibrations. At simplest, the interaction of the pyrrole ring with the NP "clamps" the N atom to the gold, deforming the ring at the N site and quenching the corresponding C-N-C mode. For this reason, the Au-N vibration in the gold-loaded micelles replaces the C-N-C vibration present in the metal-free micelles. The strong absorption at 359 cm$^{-1}$ confirms that the NPs remain inside the inverse micelles, wrapped in the polar heads of the core ligand, in all samples. We suggest that the weaker modes at 143 cm$^{-1}$ and 190 cm$^{-1}$ in the NP samples could also be associated with gold-micelle modes.

THF is water-miscible and toluene is water-insoluble. Water is present only in the gold micellar solutions in THF (blue curves in Fig. 1b and Fig. 1c) with concentrations of $c$=53 mM for charged NP and $c$=63 mM for neutral NP solutions, respectively. Two broad THz absorption features of liquid water peak at ~170 cm$^{-1}$ and ~450 cm$^{-1}$, see Fig. 1b and Fig. 1c. These features are assigned to the hindered translational (~170 cm$^{-1}$), also called intermolecular hydrogen bond stretch in the manuscript, and librational (~450 cm$^{-1}$) collective motions of hydrogen-bonded water molecules[30]. The peak absorption frequencies are red shifted by about 20 cm$^{-1}$ and 220 cm$^{-1}$ compared to bulk water[28,55], respectively.

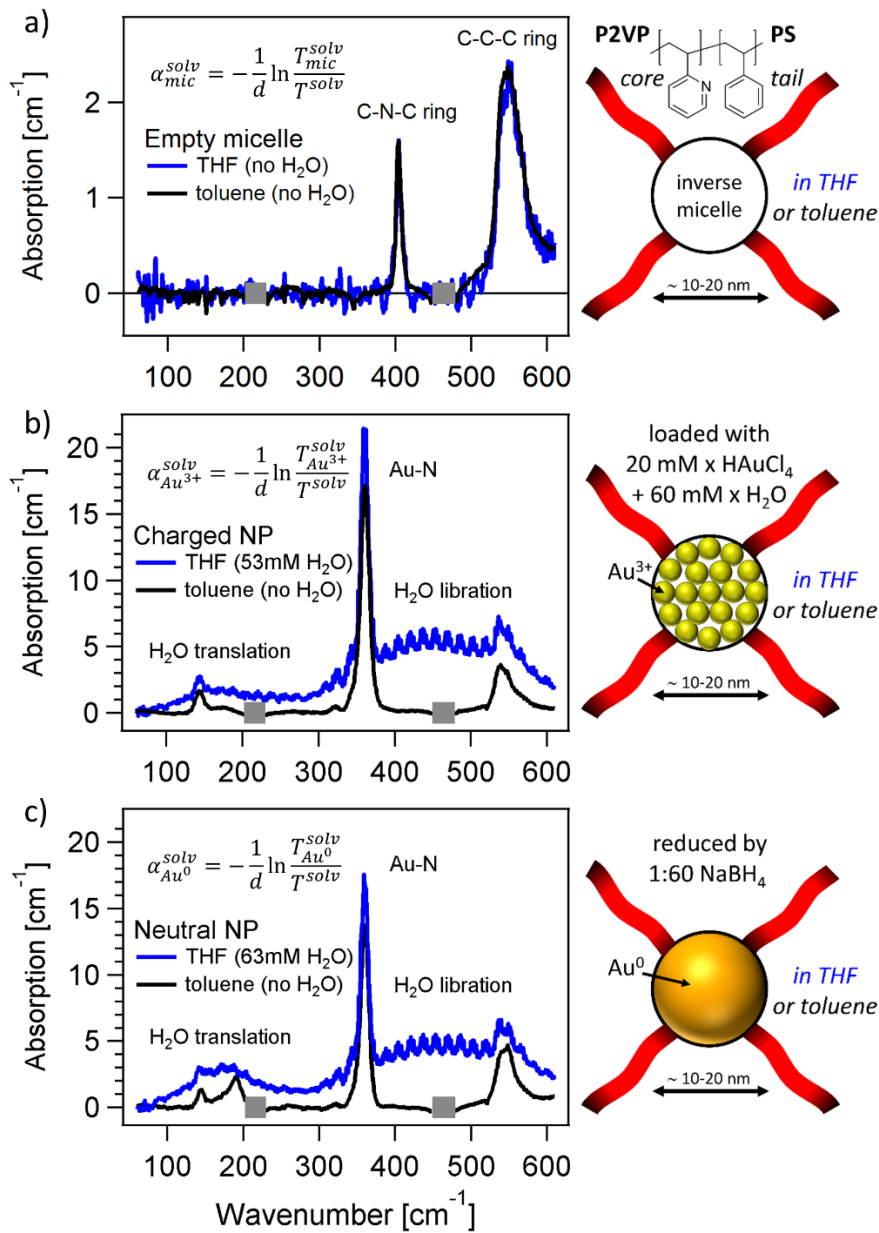

**Fig. 1**. Terahertz absorption of micelles in polar (THF, blue curves) and non-polar (toluene, black curves) solvent. The absorption coefficient of each solute $i$ (metal-free micelles, $Au^{3+}$ NP, $Au^0$ NP) is estimated to $\alpha_i^{solv} = -\frac{1}{d}\ln\frac{T_i^{solv}}{T^{solv}}$, where $d$ is the spacer thickness and $solv$ is either pure THF or toluene solvents. The micelles are either metal-free (a), loaded with charged gold nanoparticles (b), or with neutral gold NPs (c). Panels a), b), and c) share the same x axis (indicated on the bottom). The gray bands mark the position of saturated absorption peaks of pure toluene[56]. The absorption coefficients of all the toluene-based solutions is multiplied by 1.5. Vibrational features of the inverse micelles can be found at 551 cm$^{-1}$ and 404 cm$^{-1}$. The broad absorption bands of liquid water, collective translations peaking at ~170 cm$^{-1}$ and librations at ~450 cm$^{-1}$, are present in THF only (blue curves). On the right we show cartoons of the corresponding micelles: a) metal-free, b) loaded with charged NP, and c) micelles filled with $Au^0$. The red bands represent the P2VP-PS ligand.

**MIR absorption**. Behafarid et al.[35] addressed the electronic and structural properties of this kind of micellar gold nanoparticles with combined X-ray absorption spectroscopy, scanning transmission electron and atomic force microscopy, and density functional calculations. They found that, for nanoparticles of the size we have here, the electronic interactions between the polymeric ligand shell and the $Au^{3+}$ or $Au^0$ NPs are negligible and can be considered as separate, independent electronic systems. In the MIR range, no vibrations involving the heavy gold atoms are expected. In order to focus only on the changes due to charge we calculate the ratio $\frac{T_{NP}^{solv}}{T_{mic}^{solv}}$ between the transmission of gold-loaded ($T_{NP}^{solv}$) and metal-free micelle solutions ($T_{mic}^{solv}$). The raw MIR transmission measurements are reported in Supplementary Fig. 3.

Water is present only in solution with THF, and the MIR molar absorption of the water molecules surrounding differently charged NP can be estimated to $a_{wat}^{MIR} \approx \left(-\frac{1}{d}\ln\frac{T_{NP}^{THF}}{T_{mic}^{THF}}\right)/c$, where $d$ is the sample thickness and $c$ the concentration of water. Fig. 2a displays this MIR absorption of the hydration water molecules in solution with charged (yellow) and neutral gold nanoparticles (orange), together with the absorption by bulk water (blue). When compared with the MIR absorption by bulk water, the absorption linewidth is reduced for water in solution with both NP species. The magnitude of the water absorption is increased at higher frequencies for the NP solutions: the high (low) frequency components of the O-H intramolecular stretch mode absorb more (less) radiation than bulk water.

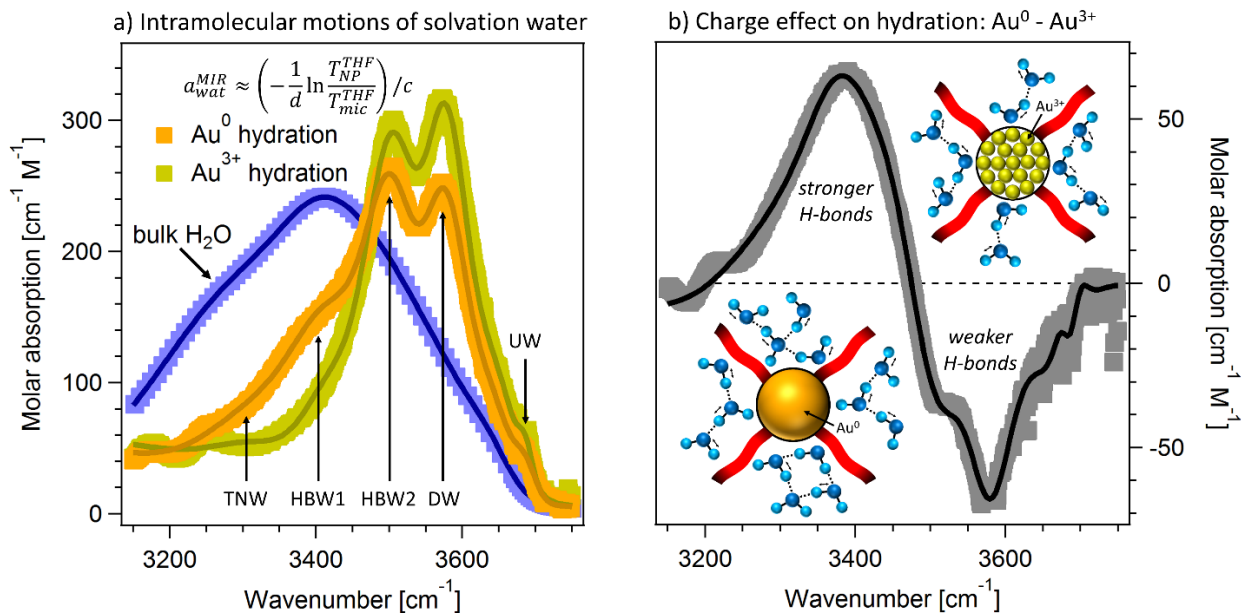

**Fig. 2**. Mid-infrared absorption of water around $Au^{3+}$ and $Au^0$ nanoparticles. The molar absorption coefficient of water hydrating each species of NP ($Au^0$ or $Au^{3+}$) is estimated to be $a_{wat}^{MIR} \approx \left(-\frac{1}{d}\ln\frac{T_{NP}^{THF}}{T_{mic}^{THF}}\right)/c$. Here $c$ is the water concentration, $d$=0.5 mm the sample thickness, $T_{mic}^{THF}$ the transmission by metal-free micelles in solution with THF, and $T_{NP}^{THF}$ the transmission of THF solutions containing gold-loaded micelles. The raw transmission data are displayed in Supplementary Fig. 3a. The water concentration is obtained

by normalizing the integral of the absorption within 3200 cm$^{-1}$ and 3700 cm$^{-1}$ to the value of bulk water (see Supplementary Fig. 3). a) Molar absorption of bulk water (blue) and of the water molecules in solution with charged (yellow) or neutral (orange) nanoparticles. The solid lines are the result of a global Gaussian fit, see Supplementary Table 1. The most relevant peaks are labeled as TNW (tetrahedral network water), HBW1 and HBW2 (hydrogen bonded water), DW (disordered water), and UW (unbound water). b) Effect of the nanoparticle charge on the intramolecular O-H stretch vibration of the water molecules. The difference of the orange and yellow curves in panel a) is shown in gray in panel b). The solid black line is the difference between the corresponding fits. The increased MIR absorption at lower wavenumbers and the lower absorption at higher wavenumbers suggests that the strength of the hydrogen bonds decreases when the nanoparticles are charged.

**Water location.** The absorption by water in the THz range has two characteristic broad peaks[30], centered at ~190 cm$^{-1}$ and 670 cm$^{-1}$. These collective water modes are absent in all micellar samples in toluene (black lines in Fig. 1). This holds both for metal-free inverse micelles (Fig. 1a) and for NP-loaded micelles (Fig. 1b,c). Given that the sensitivity of the measurement is better than $\Delta\alpha$ = 0.05 cm$^{-1}$ on the absorption axis, and considering that the absorption of pure liquid water is e.g. $\alpha$~1200 cm$^{-1}$ at v=300 cm$^{-1}$ wavenumbers[55,57], we deduce an upper limit of water in all the toluene samples of less than 0.05 cm$^{-1}$/1200 cm$^{-1}$ x 55.4 M = 2 mM. This sub-2 mM H$_2$O concentration is 30x less than the amount of water which was added before to the toluene solution to form the NPs (20 mM of HAuCl$_4$ and 60 mM of H$_2$O). When water mixes with toluene a phase separation occurs, and the denser water goes at the bottom of the container[58]. We propose that water is mostly located outside the micelle, where upon solvation the toluene phase separation occurs. We pipette the sample by soaking the tip of the syringe at the top of the solution, where small amounts of water are expected. After loading the samples into the static liquid cell for transmission spectroscopy, the water residuals further separate from toluene and settle to the bottom. The sample edges are not illuminated by the far-infrared beam and are not probed.

This interpretation is confirmed by the experimental investigation of the micellar samples in water-soluble THF. No phase separation is expected in the THF solutions containing charged as well and neutral gold NPs inside the inverse micelles (blue curves in Fig. 1b,c). We clearly observe water modes in the THz as well as in the MIR as shown in SI Fig.3. Based on these measurements, we estimate that ~60 mM H$_2$O are present in gold-loaded micelles in THF. This amount corresponds to the amount of water added before to form NPs (20 mM of HAuCl$_4$ and 60 mM of H$_2$O).

Taken together, all these results suggest that the gold metal is confined inside the inverse micelles, wrapped in the polar heads of the polymer, and that water is preferably found outside of the micelles.

**Discussion**

**Confinement.** Inverse micelles with >10 nm diameter could contain several thousand water molecules, with a corresponding molar ratio of water to surfactant[33], $\omega_0$, of more than 30. Inside large inverse micelles water display MIR and THz spectra which are very similar to pure bulk water, see e.g. Fig.2 in ref[33] and Fig.2 in ref[60]. However, the MIR (Fig.2) and THz (Fig.1, Fig.3) measurements presented here, show absorption features which are very different with respect to bulk water. Thus, as also supported by multiple other experimental results outlined in the previous paragraph, we propose that water is not primarily encapsulated inside the micelles studied here. For these reasons, to first order, we can neglect the effect of confinement in the following discussion.

**Bulk water vs. hydration water around NP**. As shown in Fig. 1, the THz absorption of the hydration water of the nanoparticles is found to be red shifted with respect to bulk water, and has narrower linewidths. In the MIR, the absorption by water in the NP solutions is blue shifted with respect to the bulk and is characterized by a reduced linewidth, see Fig. 2. These results both imply that the overall strength of the hydrogen bonding between the water molecules surrounding either charged or neutral gold nanoparticles is weaker than in bulk water[25,55,61]. We suggest that the small amount of water present in solution with the gold NP (~60 mM) is unable to form a complete 3D network and shows reduced cooperativity.

**MIR absorption of hydration water**. The interaction between THF and water and the effect of THF on the hydrogen bonding has been addressed in a recent publication by[59] M.J. Shultz and T.H. Vu. These authors prepared dry carbon tetrachloride and analyzed how the intramolecular water spectrum is affected by addition of THF. When THF is added, the O-H water spectrum develops a broad absorption feature spanning 3400-3600 $cm^{-1}$ and a sharp peak at ~3685 $cm^{-1}$. The broad absorption is assigned to an H-donor resonance associated to a complex consisting of 2x THF molecules and 1x $H_2O$, and the sharp peak to an H-bond acceptor interaction between the water oxygen atom and the α-methylene group of the ether. We measured the intramolecular O-H absorption by water in solution with gold-loaded micelles in THF (Fig. 2, Suppl. Fig. 3, Fig. 3). The MIR spectra reveal that water is, as expected, partially hydrogen-bonded with the abundant THF solvent. The orange and yellow curves in Fig.2a peak around 3500 $cm^{-1}$, and a shoulder at ~3700 $cm^{-1}$ is also present. The higher OH stretch frequencies corresponds to weakly hydrogen-bonded configurations, and lower OH stretch frequencies correspond to strongly hydrogen-bonded configurations[62]. Thus, upon charging the gold nanoparticle the hydrogen-bonded network becomes weaker since we observe a blue shift in the OH stretch region (see e.g. Fig. 2a).

The interpretation of the mid-IR spectrum of $H_2O$ is complex due to inter- and intra-molecular couplings between the OH vibrations[63], and caution is needed in assigning specific peaks in the spectra to specific water sub-populations. With these limitation in mind, here, in a first order approach, we globally fitted the absorption of bulk water, water around $Au^{3+}$, and water in solution with $Au^0$, with a sum of Gaussian peaks[59,61]. Details of the fitting procedure can be found in Supplementary Table 1. The most relevant peaks describing the absorption by water in solution with nanoparticles are indicated in Fig. 2a. The O-H mode of liquid water has been extensively studied[25,28,59,61] and can be used to probe the strength of the hydrogen bond. We associate[61] the intramolecular absorption centered at ~3321 $cm^{-1}$ to fully tetrahedrally coordinated water (tetrahedral network water, TNW). Two peaks are related to water molecules with an intermediate coordination number, either hydrogen-bonded with other water molecules[61] (hydrogen bonded water 1, HBW1, ~3427 $cm^{-1}$) or with THF molecules[59] (hydrogen bonded water 2, HBW2, ~3505 $cm^{-1}$). By comparison with previous works[59,61], we suggest that the absorption peak at ~3578 $cm^{-1}$ is due to water molecules with a lower number of hydrogen bonds (disordered water, DW), and that the shoulder at 3686 $cm^{-1}$ originates from the "dangling" OH stretch (unbound water, UW).

The low frequency peaks TNW and HBW1, associated to water molecules with higher coordination, are considerably decreased for a charged gold NP (Fig. 2a). The absorption of DW and UW is related to weakly hydrogen bonded molecules, and increases in case of $Au^{3+}$, see Fig. 2a. Thus, when compared to neutral NP, the weighted average frequency of the O-H stretch mode is blue shifted for water in solution with charged NPs. This weighted frequency amounts to $3529\ cm^{-1}$ and $3493\ cm^{-1}$ for water molecules around charged and neutral nanoparticles, respectively, see Supplementary Table 1. In Fig. 2b we show the difference between the MIR molar absorption of water in solution with $Au^0$ NP (orange in Fig. 2a) and $Au^{3+}$ NP (yellow in Fig. 2a).

The frequency of an intramolecular O-H stretch in the liquid phase is affected by the local number, strength, and direction of the hydrogen bonds[26,27,29]. For a full 3D network, a useful quantity accounting for the strength of the hydrogen bonds in liquid water is the order parameter $q$. This parameter measures the extent to which a molecule and its nearest neighbors adopt a tetrahedral configuration[25,64]: for a perfect tetrahedron $q = 1$, for an ideal gas $q = 0$. Morawietz et al.[25] recently performed a combined experimental and theoretical study and quantified how the O-H stretch frequency correlates with the mean local tetrahedral order of a water molecule and, in this sense, with the overall strength of the hydrogen bond network. From the weighted average frequency of the whole O-H stretch mode of the water molecules in solution with $Au^{3+}$ and $Au^0$ NP (see Supplementary Table 1), we estimate[25] that the local tetrahedral order parameter is $q \approx 0.65$ for $H_2O$ in solution with $Au^0$, and $q \approx 0.5$ for $H_2O$ hydrating $Au^{3+}$. We stress that this estimation is strictly valid for a 3D hydrogen bond water network. However, we are aware that we have few water molecules (~60 mM), and we do not expect to form a full 3D network.

**THz absorption of hydration water**. The THz absorption of metal-free and gold-loaded micelles can be affected by vibrations at the interface between the ligand and the NP. For example, the motion of the 2-vinylpyridine pyrrole ring[51–53] and of the gold atoms at the surface of the nanoparticle[54] are centered below 600 cm$^{-1}$. As detailed in Fig. 1, in order to benchmark possible low-frequency gold motions we measured the THz absorption of the micellar solutions also in a solvent where no water is present (toluene). The THz molar absorption coefficient of water surrounding each species of NP can be approximated to $a_{wat}^{THz} \approx (\alpha_{NP}^{THF} - \alpha_{NP}^{tol})/c$. This corresponds to the difference between the blue and the black curves in Fig. 1b for $Au^{3+}$ NP, and between the blue and black curves in Fig. 1c for neutral NP, each normalized on the corresponding water concentration. The molar absorption coefficient of the hydration water in the THz range, $a_{wat}^{THz}$, is shown in Fig. 3a: yellow for $Au^{3+}$ and orange for $Au^0$ solutions. For comparison, the THz absorption of pure water is shown in blue. It is evident that the THz linewidths are reduced for both solvated $Au^{3+}$ NP as well as solvated $Au^0$ NP. The low frequency modes are red shifted for both NP solutions when compared to bulk water. The translational mode peaks at ~170 cm$^{-1}$ for water in solution with NP, and at ~190 cm$^{-1}$ for bulk water[25,28,55] (red shift of about 20 cm$^{-1}$). The librational absorption is found to be considerably red shifted with a maximum at ~450 cm$^{-1}$ for water around nanoparticles, compared to ~670 cm$^{-1}$ in bulk water[25,28,55].

In order to quantify the different vibrational response of the hydrogen-bonded water network in solution with charged NP (yellow curve in Fig. 3a) and neutral NP (orange in Fig. 3a), we fitted the THz absorption with two damped harmonic oscillators[65]. The details of the fitting procedure are included in Supplementary Table 2. Within the fit errors, the librational and translational bands of the hydration water of charged and neutral NP have the same central frequency and same width, but different amplitude. We estimated the absorption intensity by multiplying the peak amplitude and the width, and found that positive-charged nanoparticles decrease the intensity of the collective stretch of water by (-19±2)% and increase the intensity of the librational mode by (+32±1)%. These results are consistent with the results of Schienbein et al.[65], who recently showed that a metal cation lead to a reduced number of hydrogen bonds in its first hydration shell, decreasing the effective amplitude of the translational mode while increasing the librational intensity. However, the intensity is also affected by polarizability effects, which are particularly important for collective water motions in the THz range[30,65].

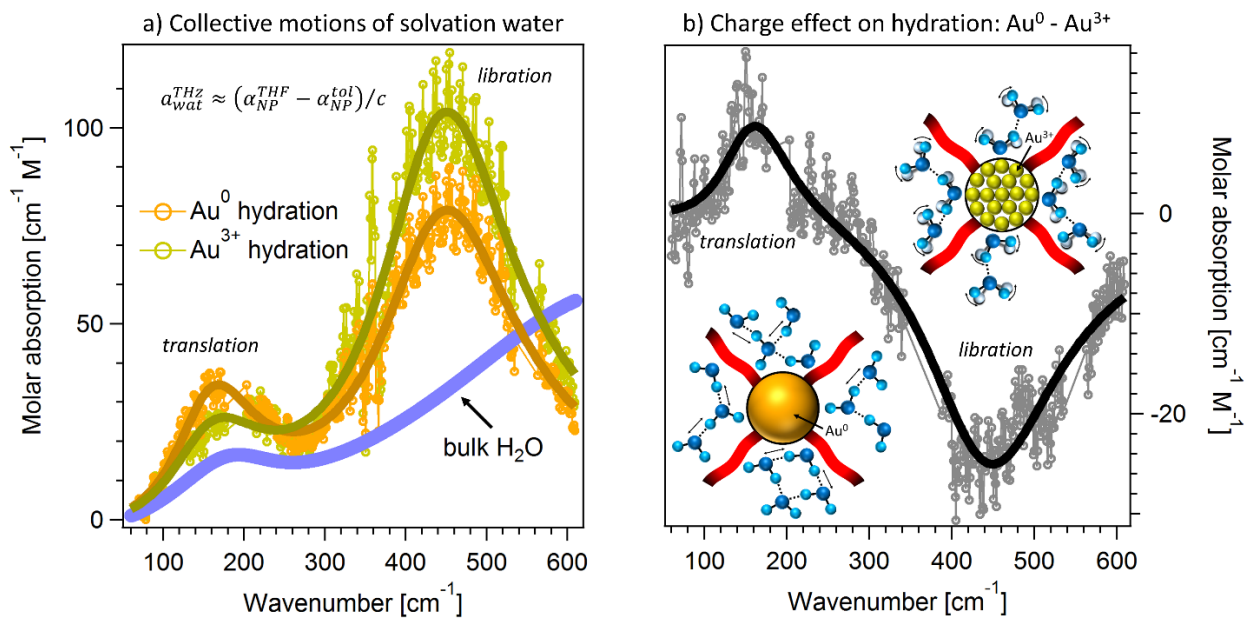

**Fig. 3**. Terahertz absorption of water around nanoparticles with different charge. The molar absorption coefficient of water around each species of NP ($Au^0$ or $Au^{3+}$) is estimated to $a_{wat}^{THz} \approx (\alpha_{NP}^{THF} - \alpha_{NP}^{tol})/c$, with $c$ water concentration, and $\alpha$ the absorption coefficients shown in Fig. 1. a) Molar absorption of bulk water (blue) and of the water molecules solvating charged (yellow) or neutral (orange) nanoparticles. An offset of -5 cm$^{-1}$ M$^{-1}$ is added to the bulk water absorption for display purposes (blue curve). The solid lines are the result of fits detailed in Supplementary Table 2. b) Effect of the nanoparticles charges on the intermolecular absorption by liquid water in the hydration layer. The difference between the orange and yellow curves in panel a) is shown in gray in panel b). The solid black line is the difference between the fits in panel a). The intensity of the translational band increases, while the intensity of the libration mode decreases.

From the MIR measurements, we estimated that the local tetrahedral order parameter of water in solution with the NP is reduced respect to bulk water ($q \approx 0.65$ for $Au^0$, $q \approx 0.5$ for $Au^{3+}$). Reducing the coordination in the 3D network implies that the intermolecular modes, such as the translational mode and the librational mode, should be red shifted compared to bulk water[25,55]. The librational mode of the hydration water around NP species is centered at 440 cm$^{-1}$, see Fig. 3 and Supplementary Table 2. We propose that the enormous red shift compared to bulk water is due to the decreased dimensionality of the water network around the nanoparticle, implying that the cooperativity is lower than for a full 3D network.

**Conclusions**

We have shown how water interacts with the charge of ~10 nm sized solutes. We recorded the mid-infrared and terahertz absorption spectra of water surrounding gold-loaded micelles and found that it differs from bulk water. In the O-H stretch region we observed an overall blue shift for the hydration water around nanoparticles. Both intermolecular modes in the THz range are found to be red shifted. These observations indicate that the intermolecular hydrogen bond is weakened compared to bulk water.

Based upon the MIR measurements we were able to estimate the order parameter $q$, which quantifies how a water molecule and its nearest neighbors adopt a tetrahedral configuration in 3D[25,64]. The $q$ parameter is $\approx 0.65$ for H$_2$O in solution with Au$^0$, and $q \approx 0.5$ for H$_2$O hydrating Au$^{3+}$. These values reveal that the water network hydrating both neutral as well as charged gold nanoparticles is not a full 3D network. The peak frequency of the librational mode of water in the THz is at 670 cm$^{-1}$. In comparison, for the hydration water of the nanoparticle we find a peak at 440 cm$^{-1}$, see Fig. 3. This confirms a weakening of the hydrogen bond network associated with a reduction in cooperativity, for which the 3D network cannot develop completely.

Further insight is gained by comparing the absorption of the hydration water of positive and neutral gold nanoparticles. For positively charged NP, the center of the integrated absorption in the O-H stretch band is found to be blue shifted, suggesting that the hydrogen bonds are weaker around charged nanoparticles compared to those around neutral NP. The translational and librational THz modes of the hydration water around charged and neutral NP have similar central frequencies, within their error bars (Supplementary Table 2).

Water at a dielectric interface, such as between vacuum and bulk[29] or around an ion[66–69], experiences a different local surface potential. Based on the fact that the THz librational mode shifts to higher frequency, Tong *et al*.[29] suggested that the potential for the libration of water is steeper at the interface of water and air compared to bulk water. Conversely, the orientational dynamics of the hydration water can be either accelerated or slowed according to the specific anion[67]. Recent results[66] indicated that both sodium and potassium cations could accelerate the non-diffusive water rotations. We speculate that the unchanged central frequencies of the THz bands for solvated Au$^{3+}$ compared to Au$^0$ NP solutions might originate from a combination of a stiffer local potential[31] experienced by the water at the interface with the charged gold[29], resulting in a blue shift of the librational mode, and from the reduced strength of the hydrogen bond, which would cause a red shift of this mode.

This work provides insight into how water interacts with charged versus non-charged nanoparticles, and might help to improve the electro-catalytic applications of gold nanoparticles[1,2,11–20,3,21,4–10].

**Supporting Information Description**

**Amount of water in the metal-free micelles in THF.**

**Supplementary Fig.1**. Visible absorption spectra of the nanoparticles solutions.

**Supplementary Fig.2**. AFM images of the gold NPs.

**Supplementary Fig.3.** Mid-infrared measurements of the Au$^0$ and Au$^{3+}$ solutions.

**Supplementary Table 1**. Fit parameters of the MIR measurements.

**Supplementary Table 2.** Fit parameters of the THz measurements.

**Acknowledgement**

We acknowledge T. Lux for experimental support and P. B. Petersen for useful comments. This work received funding from the ERC Advanced Grant 695437, and is part of the excellence cluster RESOLV funded by the Deutsche Forschungsgemeinschaft (DFG, German Research Foundation) under Germanýs Excellence Strategy – EXC-2033 – Projektnummer 390677874.